\documentclass[12pt]{iopart}

\usepackage{citesort,latexsym,ifthen,graphics,color}
\newcommand{\nc}{\newcommand}
\newcommand{\nt}{\newtheorem}

\newcounter{Diagrams}
\Alph{Diagrams}
\newcounter{cancellation}
\addtocounter{cancellation}{1}

\nt{D}{}[Diagrams]
\nt{DC}[D]{\{}
\nt{cancel}{Cancellation}

\nc{\QED}{QED}
\nc{\ERG}{ERG}
\nc{\ERGs}{ERGs}
\nc{\ibid}{\emph{ibid.}}
\nc{\viz}{viz}
\nc{\eg}{e.g.\ }
\nc{\ie}{i.e.\ }
\nc{\cf}{cf.\ }
\nc{\etc}{etc.\ }
\nc{\rhs}{right-hand side}
\nc{\lhs}{left-hand side}
\nc{\wrt}{with respect to}
\nc{\aka}{a.k.a.\ }
\nc{\sic}{sic}
\nc{\QFT}{QFT}
\nc{\BS}{\Lambda_0}
\nc{\ES}{\Lambda}
\nc{\PV}{PV}
\nc{\ala}{\emph{\`a la}}
\nc{\role}{role}
\nc{\vev}{\emph{vev}}

\nc{\be}{\begin{equation}}
\nc{\ee}{\end{equation}}
\nc{\bea}{\begin{eqnarray}}
\nc{\eea}{\end{eqnarray}}
\nc{\beas}{\begin{eqnarray*}}
\nc{\eeas}{\end{eqnarray*}}

\nc{\bcf}{\begin{center}\begin{figure}}
\nc{\ecf}{\end{figure}\end{center}}

\nc{\bct}{\begin{center}\begin{table}}
\nc{\ect}{\end{table}\end{center}}

\nc{\ds}{\displaystyle}
\nc{\sst}{\scriptstlye}

\nc{\eq}[1]{(\ref{#1})}
\def\eq#1{(\ref{#1})}

\def\sec#1{section~\ref{sec:#1}}

\def\fig#1{figure~\ref{fig:#1}}

\nc{\str}{\mathrm{str}\,}
\nc{\diag}{\mathrm{diag}}
\nc{\strI}{\mathrm{str} \!\! \int \!\! d^D \! x\,}
\nc{\ODInt}[1]{\int \!\! d#1}
\nc{\ODIntL}[3]{\int_{#2}^{#3} \!\! d#1}
\nc{\Int}[1]{\int \!\! d^D \! #1 \,}
\nc{\IntB}[1]{\int \!\! \frac{d^D \! #1}{(2\pi)^D} \,}
\nc{\DoubleInt}{\int\!\!\!\!\int}
\nc{\volume}[1]{d^D \! #1 \,}
\nc{\ve}[1]{d^D \! #1}
\nc{\AngVol}[1]{\not{\!\Omega_#1}}
\nc{\PowAngVol}[2]{\not{\! \! \Omega_#1^#2}}

\nc{\af}[1]{\bar{#1}}

\nc{\desl}{\partial \hspace{-.51em}/ }
\nc{\asl}{A \hspace{-.48em}/ }
\nc{\psl}{p \hspace{-.51em}/ }
\nc{\ksl}{k \hspace{-.51em}/ }
\nc{\Osl}{\not{\! \Omega}} 

\nc{\A}{\mathcal{A}}
\nc{\C}{\mathcal{C}}
\nc{\GRk}{\rhd}
\nc{\GRkpr}{\ensuremath{>}}

\nc{\gam}{\gamma}
\nc{\Lam}{\Lambda}
\nc{\chibar}{\bar{\chi}}
\nc{\psibar}{\bar{\psi}}
\nc{\Phibar}{\bar{\Phi}}

\nc{\demu}{\partial_{\mu}}
\nc{\denu}{\partial_{\nu}}

\nc{\nCr}[2]{\ensuremath{\left.^{#1}\right. \! C_{#2}\ }}
\nc{\norm}{\ensuremath{\Upsilon}}
\nc{\LDs}[1]{\mathcal{D}_{#1}}

\nc{\kernel}[1]{\!\cdot #1\!\cdot\!}
\nc{\covkernel}[1]{\{ #1 \}} 

\nc{\bigdot}[1]{\stackrel{\bullet}{#1}}
\nc{\DEP}{\bigdot{\Delta}}
\nc{\DGRkpr}{\stackrel{\bullet}{\GRkpr}}

\nc{\dd}{\dot{\Delta}}
\nc{\ci}{c^{-1}}
\nc{\hS}{\hat{S}}
\nc{\wrn}[2]{\gamma^{(#1)}_{#2}}

\nc{\comm}[2]{\left[ #1,#2 \right]}

\nc{\one}{\ensuremath{1\! \mathrm{l}}}
\nc{\der}[2]{\ensuremath{\frac{d #1}{d #2}}}
\nc{\pder}[2]{\ensuremath{\frac{\partial #1}{\partial #2}}}
\nc{\fder}[2]{\ensuremath{\frac{\delta #1}{\delta #2}}}
\nc{\order}[1]{\mathcal{O}\left( #1 \right)}
\nc{\Op}[1]{\mathcal{O}(p^{#1})}
\nc{\hf}{\frac{1}{2}}

\nc{\flow}{\ensuremath{\Lambda \partial_\Lambda }}
\nc{\flowConstAl}{\ensuremath{\Lambda \partial_\Lambda|_\alpha }}
\nc{\dec}[3][0]{\ensuremath{\left[ #2 \hspace{#1in} \right]^{#3}}}
\nc{\TLTP}[1]{S_{0 \mu \,\nu}^{\ AA}(#1)}
\nc{\OLDs}{\mathcal{D}_1}

\nc{\DAD}{\scriptscriptstyle \odot}

\nc{\DiagDot}{\scriptstyle \bullet}
\nc{\DummyKernel}{\ensuremath{\stackrel{\bullet}{\mbox{\rule{1cm}{.2mm}}}}}
\nc{\DecKernel}{\ensuremath{\stackrel{\circ}{\mbox{\rule{1cm}{.2mm}}}}}

\nc{\inte}{\!\int\!}
\nc{\dga}{\ensuremath{{\cal D}_{\Gamma}}}

\nc{\nothing}{}	

\newlength{\LabLength}
\newlength{\VertexWidth}

\nc{\Vertex}[1]{
\ensuremath{
	\begin{array}{c}
	\settowidth{\VertexWidth}{$#1$}
	\setlength{\unitlength}{1.2\VertexWidth}
	\begin{picture}(1,1)(-0.5,-0.5)
		\put(0,0){\circle{1}}
		\put(-0.4,-0.1){$#1$}
	\end{picture}
	\end{array}
}
}

\newcommand{\SumVertex}{
	\Vertex{n_s, j}
}

\newcommand{\TopVertex}{
	\Vertex{v^{j_+;R}}
}

\nc{\cd}[1]{\ensuremath{\begin{array}{c}\input{pstex/#1.pstex_t} \end{array}}} 
\nc{\cdtemp}[1]{\ensuremath{\begin{array}{c}\input{#1.pstex_t} \end{array}}} 
\nc{\cdeps}[1]{\ensuremath{\begin{array}{c}\includegraphics{eps/#1.eps} \end{array}}}

\nc{\sco}[3][0]{
	\begin{array}{c}
		#2 \\[#1ex]
		#3
\end{array}
}

\nc{\scd}[3][0]{
	\sco[#1]{\cd{#2}}{\cd{#3}}
}

\nc{\LD}[1]{
	\settowidth{\LabLength}{\scriptsize \textbf{\ref{#1}}}
	\addtolength{\LabLength}{0.8em}
	\begin{minipage}{\LabLength}
		\scriptsize
		\begin{D}\label{#1}\end{D}
	\end{minipage}
}

\nc{\LO}[3][1]{
	\begin{array}{c}
		\LD{#3}
	\\[#1ex]
		#2
	\end{array}
}

\nc{\LDi}[3][1]{\LO[#1]{\cd{#2}}{#3}}

\nc{\jhep}[3]{\emph{JHEP} #1 (#2) #3}
\nc{\NuclPhys}[4]{\emph{Nucl.\ Phys.\ }\textbf{#1 #2} (#3) #4}
\nc{\PhysRev}[4]{\emph{Phys.\ Rev.\ }\textbf{#1 #2} (#3) #4}
\nc{\IntJModPhys}[4]{\emph{Int.\ J.\ Mod.\ Phys.\ }\textbf{#1 #2} (#3) #4}
\nc{\PhysRep}[4]{\emph{Phys.\ Rep.\ }\textbf{#1 #2} (#3) #4}
\nc{\PhysRept}[3]{\emph{Phys.\ Rept.\ }\textbf{#1} (#2) #3}
\nc{\TheorMathPhys}[3]{\emph{Theor.\ Math.\ Phys.\ }\textbf{#1} (#2) #3}
\nc{\ModPhysLett}[4]{\emph{Mod.\ Phys.\ Lett.\ }\textbf{#1 #2} (#3) #4}
\nc{\AnnPhys}[3]{\emph{Ann.\ Phys.\ }\textbf{#1} (#2) #3}
\nc{\PhysLett}[4]{\emph{Phys.\ Lett.\ }\textbf{#1 #2} (#3) #4}
\nc{\ProgTheorPhys}[3]{\emph{Prog.\ Theor.\ Phys.\ }\textbf{#1} (#2) #3}
\nc{\Acta}[3]{\emph{Acta Phys.\ Slov.\ }\textbf{#1} (#2) #3}
\nc{\EurPhysJ}[3]{\emph{Eur.\ Phys.\ J.\ }\textbf{#1} (#2) #3}
\nc{\CEurJPhys}[3]{\emph{Central Eur.\ J.\ Phys.\ }\textbf{#1} (#2) #3}
\nc{\ZPhys}[4]{\emph{Z.\ Phys.\ } \textbf{#1 #2} (#3) #4}
\nc{\PhysRevLett}[3]{\emph{Phys.\ Rev.\ Lett.\ } \textbf{#1} (#2) #3}
\nc{\ClassQuantGrav}[3]{\emph{Class.\ Quant.\ Grav.\ }\textbf{#1} (#2) #3}
\nc{\arxiv}[1]{[arXiv:#1]}
\nc{\hepth}[1]{hep-th/#1}
\nc{\hepph}[1]{hep-ph/#1}
\nc{\heplat}[1]{hep-lat/#1}

\nc{\http}[1]{http://#1}

\begin{document}

\title{Scheme Independence to all Loops}

\author{
	Oliver J.~Rosten
}

\address{School of Physics and Astronomy,  University of Southampton,
	Highfield, Southampton SO17 1BJ, U.K.}

\ead{O.J.Rosten@soton.ac.uk}

\begin{abstract}
	The immense freedom in the construction of 
	Exact Renormalization Groups means that 
	the many non-universal details of the formalism 
	need never be exactly specified, instead
	satisfying only general constraints. 
	In the context of a manifestly gauge invariant
	Exact Renormalization Group for $SU(N)$ Yang-Mills,
	we outline a proof that, to all orders in perturbation
	theory, all explicit dependence of $\beta$ function
	coefficients on both the seed action and details of
	the covariantization cancels out. Further, we speculate
	that, within the infinite number of renormalization
	schemes implicit within our approach, the perturbative
	$\beta$ function depends only on the universal details
	of the setup, to all orders.	
\end{abstract}

\vspace{-60ex}
\hfill SHEP 05-34
\vspace{61ex}

\maketitle

\section{Introduction and Conclusions}

The Exact Renormalization Group (ERG) provides
an extremely flexible framework for dealing
with Quantum Field Theory, as a direct consequence
of the immense freedom in its
actual construction~\cite{TRM+JL,mgierg1}. 
Remarkably, this
freedom can be exploited to construct
ERGs for $SU(N)$ Yang-Mills theory in which the gauge
invariance is \emph{manifestly} 
preserved~\cite{mgierg1,mgierg2,Primer,Quarks,Thesis,aprop,ym,ym1,ym2}, 
with the realization of a gauge invariant
cutoff being achieved by embedding the physical
theory, carried by the field $A^1_\mu$, in a spontaneously broken $SU(N|N)$ gauge
theory~\cite{sunn}.
In addition to the
coupling, $g$, of the physical gauge field, there is a second running,
dimensionless coupling, $g_2$, associated with one of the unphysical
regulator fields, $A^2_\mu$. As a consequence of  $g$ and $g_2$
renormalizing separately, it is convenient to 
further specialize
to those ERGs which treat $A^1_\mu$ and $A^2_\mu$ asymmetrically,
in the broken phase~\cite{Thesis,mgierg1}. The resulting formalism has been
used to successfully compute the $SU(N)$ Yang-Mills two-loop
$\beta$ function, $\beta_2$~\cite{Thesis,mgierg1,mgierg2,Primer}, without
fixing the gauge.

Despite the restrictions we choose to impose on the ERG,
there are still an infinite number with which we can work,
corresponding to the residual freedom in the precise details
of the construction; and this residual freedom, too, can 
be turned to our advantage. The differences between the
admissible ERGs amount to non-universal details
which, whilst encoding the ultra-violet regularization,
parameterize the continuum notion of a general Kadanoff
blocking procedure~\cite{TRM+JL,mgierg1}.
Our philosophy is to leave the non-universal 
details largely 
unspecified, defining them implicitly where necessary,
to ensure that the flow equation is well behaved. 
As a direct consequence, 
the computation of universal quantities becomes particularly
efficient. A universal quantity cannot depend
on non-universal details; by leaving these details essentially
unspecified, their cancellation
becomes so constrained that
it can be performed diagrammatically. In other words,
the diagrammatics represent the natural way of encoding 
the non-universalities inherent in the flow equation.

These observations underpin the 
computation of the universal coefficient
$\beta_2$, in which diagrammatic techniques
are used to iteratively remove non-universal
contributions. 
To elucidate  this procedure, we
enumerate the various sources of
non-universality. 
First, there is the precise
form of the cutoff functions. Secondly, there are the
details of the covariantization of the cutoff. 
Lastly,
there is the `seed action', 
$\hat{S}$~\cite{mgierg1,mgierg2,Thesis,scalar1,scalar2,aprop,giqed}:
a functional which
respects the same symmetries as the Wilsonian effective 
action, $S$, and has the same structure.  However, whereas our
aim is to solve the flow equation for $S$, $\hS$ acts as an input.

The reduction of $\beta_2$
to an expression
manifestly independent of non-universal details
is a two-step procedure.
First, all explicit dependence
on the seed action and
details of the covariantization is removed.
Secondly,
all implicit dependence
on these objects and all
dependence on the shapes of the cutoff
profiles is either cancelled
or shown to vanish in an appropriate limit~\cite{mgierg2,Thesis}.

As recognized in~\cite{Thesis,Primer}, the first
diagrammatic step can in fact be employed at any number
of loops. This is, in some
respects, quite surprising. Between certain classes of
renormalization schemes, one expects agreement only between
the first two $\beta$ function coefficients~\cite{aprop,Weinberg}.
Thus, there is no \emph{a priori} reason to expect
that, within the infinite number of renormalization
schemes implicitly defined by our approach, entire sources
of non-universality will cancel out in the computation of $\beta$ function
coefficients, to all orders in perturbation theory.

It is now natural to speculate whether
the second diagrammatic step---which, like
the first, is algorithmic---can be employed at
any number of loops.  There are encouraging indications
that this is indeed the case~\cite{WIP}. Hence, we are led to speculate whether
the $\beta$ function coefficients computed in this approach
could depend only on the universal details of the setup,
to all orders in perturbation theory. It should be emphasised
that this is not a speculation
that the $\beta$ function is universal in a general
sense, since agreement would
certainly not be expected between this ERG scheme and, say, dimensional
regularization for $\beta_{n>2}$. Rather, the $\beta$ function
may be universal in a restricted sense:
if one were to imagine the complete space of renormalization
schemes, then the schemes defined by our approach would
correspond to a  `flat direction'\footnote{I would like to thank Daniel Litim
for pointing out this interpretation.} in the sense that,
when traversed, the $\beta$-function 
coefficients would remain unchanged, to all orders
in perturbation theory. 

Irrespective of whether or not universality in this restricted
sense is found, 
the reduction of $\beta$ function coefficients to a form
independent of the explicit details of the seed action
and details of the covariantization is of interest in itself.
Furthermore, we note that since this reduction is a feature
of the structure of the flow equation itself, rather than
some specific feature relating to the non-Abelian gauge 
invariance, this analysis should be trivially extendable
to scalar field theory and QED.

At this point, it is worth noting that there have been
many other ERG computations of one and two loop $\beta$
functions, in a variety of different theories and for
a variety of different 
reasons~\cite{scalar1,scalar2,Bonini1,Bonini2,Morris:1999ba,Papenbrock:1995-b2,Pernici:1998-b2,Kopietz:2001-b2,Zappala:2002-b2}. 
The emphasis
of this paper  is complete
scheme independence, inspired by
the algorithmic removal of non-universal elements,
order by order in perturbation theory. 
Indeed, this procedure 
allows us to derive an extremely compact
diagrammatic formula for arbitrary $\beta_n$
in terms of a set of objects (about which 
we will be more precise shortly) with no explicit
dependence on the seed action or details
of the covariantization.

The discovery of this formula is an
extremely important step in promoting the
ERG advocated by this paper into
a practical computational scheme,
as it represents a radical simplification
of loop calculations.
Since its inspiration
comes simply from the idea that
non-universal objects must cancel
out in the computation of universal
quantities, it is reasonable to hope
that similar simplifications
will be found in more general computations.
Specifically, for Yang-Mills theory, 
we plan to compute expectation values
of gauge invariant operators~\cite{WIP}.
An exciting possibility is that this could
guide us to
a more direct framework for performing 
manifestly gauge invariant computations, 
where objects such as the seed action would operate
entirely in the background. Indeed, it may be able to go
further and use the current scheme as inspiration for a 
manifestly gauge invariant formalism entirely independent
of the ERG, in which universality is transparent, from 
the start. We leave the investigation
of these issues for the future~\cite{WIP}.

At the heart of the diagrammatic techniques is the
`effective propagator relation'~\cite{aprop}. It is
technically useful 
to set the two-point, tree level 
seed action vertices equal to 
their Wilsonian effective action counterparts.
In turn,
this ensures that 
for each independent two-point, tree level vertex (that cannot be consistently
set to zero~\cite{Thesis}) there exists
an `effective propagator', denoted by $\Delta$, 
which is the inverse of the
given vertex, up to a `gauge remainder'.
This remainder term appears as 
a consequence of the manifest gauge invariance: the effective
propagators are inverses of the two-point, tree level vertices
only in the transverse space. The effective propagator
relation is of such central importance since it allows
the simplification of any diagram in which a two-point
vertex attaches to an effective propagator. The resulting
contributions are involved in direct
diagrammatic cancellations, up to remainders
which can themselves be processed, diagrammatically.

We now review the iterative diagrammatic procedure employed
in the first phase of the calculation of $\beta_2$~\cite{Primer,Thesis},
as a precursor to discussing the computation of arbitrary $\beta_n$.
To begin, the flow equation is used to
compute the flow of the two-point vertex corresponding
to the physical $SU(N)$ gauge field, which we suppose
carries momentum $p$. To obtain a
solvable equation for $\beta_2$,
we specialize
to the appropriate loop order and work at $\Op{2}$;
this latter step constrains the equation by allowing
the renormalization condition for the physical coupling
$g(\Lambda)$ to feed in.

We now recognize that certain diagrams generated
by the flow comprise exclusively Wilsonian
effective action vertices joined together
by $\DEP$ where, having defined
\be
\label{eq:alpha}
	\alpha \equiv \frac{g_2^2}{g^2},
\ee
we define
\be
\label{eq:dot}
	\bigdot{X} \equiv -\flowConstAl X;
\ee
$\flow$ being the generator of the \ERG\ flow. 
The manipulable diagrams are processed by moving
$\flowConstAl$ from the effective propagator
to strike the diagram as a whole, minus correction
terms in which $\flowConstAl$ strikes the vertices.
The former terms are called $\Lambda$-derivative terms;
the latter terms can be processed using
the flow equation and the resulting set of
diagrams simplified and further processed, using a set of
diagrammatic identities, of which the effective propagator
relation is one~\cite{Primer}. 
At this point, we are able to identify
cancellations of non-universal contributions.
Iterating the diagrammatic procedure,
the expression for $\beta_2$ ultimately
reduces to the following sets of diagrams:
\begin{enumerate}
	\item $\Lambda$-derivative terms;

	\item `$\alpha$-terms', consisting of diagrams
			containing a component 
			struck by $\partial / \partial \alpha$;

	\item `$\Op{2}$-terms', which contain an $\Op{2}$ stub
			\ie a diagrammatic component which is manifestly
			$\Op{2}$.
\end{enumerate}

In turn, the $\Op{2}$ terms can be diagrammatically
manipulated, thereby reducing the entire expression
for $\beta_2$ to a set of $\Lambda$-derivative terms
and a set of $\alpha$-terms. All these diagrams
contain only Wilsonian effective action vertices,
effective propagators and (components of) gauge remainders.

At arbitrary loop order, precisely the same procedure can
be employed~\cite{Thesis}. However, if the answer is
known, it is much more efficient to essentially take this as
the starting point and prove it to be true. The precise strategy
is to construct a set of terms, $\LDs{n}$, including $\beta_n$, which 
vanishes at $\Op{2}$.  Then, at $\Op{2}$, $\LDs{n}$ can be rearranged
to give a compact 
expression  for $\beta_n$ in terms of just 
Wilsonian effective action vertices, effective propagators
and (components of) gauge remainders.
 After giving the diagrammatic form
for the flow equation in \sec{flow}, in \sec{beta-n}
we partially construct $\LDs{n}$
and sketch the proof that it does
indeed vanish, as required (both the full expression for $\LDs{n}$
and the complete proof that it vanishes at $\Op{2}$
will be given in~\cite{InPrep}).

\section{The Flow Equation}
\label{sec:flow}

The diagrammatic representation of the flow
equation is shown in \fig{Flow}~\cite{Thesis,mgierg1}.
\bcf[h]
	\beas
	\ds
	-\flow 
	\dec{
		\cd{Vertex-S}
	}{\{f\}}
	& = & a_0[S,\Sigma_g]^{\{f\}} - a_1[\Sigma_g]^{\{f\}}
\\
	& = &
	\ds
	\frac{1}{2}
	\dec{
		\cd{Dumbbell-S-Sigma_g} - \cd{Padlock-Sigma_g} 
	}{\{f\}}
	\eeas
\caption{The diagrammatic form of the flow equation.}
\label{fig:Flow}
\ecf

The \lhs\ depicts the flow of all independent Wilsonian effective action
vertex \emph{coefficient functions}, 
which correspond to the set of broken phase
fields, $\{f\}$. 
Each coefficient function has associated
with it an implied supertrace structure (and symmetry factor which,
as one would want, does not appear in the diagrammatics), as expected
from the embedding of the physical gauge field into a spontaneously
broken $SU(N|N)$ theory~\cite{sunn}.
For example,
\[
	\dec{
		\cd{Vertex-S}
	}{A^1 A^1 A^1 A^1}
\]
represents both the coefficient functions $S^{A^1 A^1 A^1 A^1}$ and
$S^{A^1 A^1, A^1 A^1}$ which, respectively, are associated with the
supertrace structures $\str A^1 A^1 A^1 A^1$ and $\str A^1 A^1 \str A^1 A^1$
(there are no further coefficient functions / supertrace structures associated with
the vertex since $\str A^1 = 0$).

The dumbbell-like diagram on the \rhs\ of figure~\ref{fig:Flow}
is formed by the bilinear functional $a_0[S,\Sigma_g]$,
whereas the padlock-like diagram is formed by $a_1[\Sigma_g]$.
($a_1$ can also generate a diagram in which the
kernel `bites its own tail'~\cite{ym}. Such diagrams are improperly UV
regularized and can be removed by an appropriate constraint
on the covariantization~\cite{aprop,mgierg1}.)
Both diagrams comprise two different components. The lobes
represent vertices of action functionals,
where $\Sigma_g \equiv g^2S - 2 \hat{S}$. 
The object attaching
to the various lobes, \DummyKernel,  is
the sum over vertices of the covariantised \ERG\ kernels~\cite{ym1,aprop}
and, like the action vertices, can be decorated by fields belonging to $\{f\}$.
The fields of the action vertex (vertices) to which the vertices of the kernels attach
act as labels for the \ERG\ kernels~\cite{Thesis,mgierg1}.
We loosely refer to both individual and summed over 
vertices of the kernels simply as a kernel. 
The rule for decorating the diagrams on
the \rhs\ is simple: the set of fields, $\{f\}$, are distributed in 
all independent ways between the component objects of each diagram.
We will see an example of this in the next section.

\subsection{The Weak Coupling Expansion}

In the perturbative domain, we have the following
weak coupling expansions~\cite{Thesis,mgierg1}.
The Wilsonian effective action is given by
\be
	S = \sum_{i=0}^\infty \left( g^2 \right)^{i-1} S_i = \frac{1}{g^2}S_0 + S_1 + \cdots,
\label{eq:Weak-S}
\ee
where $S_0$ is the classical effective action and the $S_{i>0}$
the $i$th-loop corrections. The seed action has a similar expansion:
\be
	\hat{S} = \sum_{i=0}^\infty  g^{2i}\hat{S}_i.
\label{eq:Weak-hS}
\ee
The $\beta$ function is defined, as usual, to be
\be
\label{eq:beta}
	\beta \equiv \flow g  = \sum_{i=1}^\infty  g^{2i+1} \beta_i(\alpha)
\ee
where we note that $\beta_{n>1}$ are expected to depend on $\alpha$.
Indeed, this is true even at two-loops, where the universal
coefficient is recovered only in the limit that 
$\alpha \rightarrow 0$~\cite{mgierg1,mgierg2,Thesis}. Should it be possible
to introduce a notion of a universal $\beta$ function, in the
sense outlined earlier, this would presumably require that
$\alpha$ be tuned to zero, order by order. The flow
of $\alpha$ itself has the following expansion:
\be
	\gamma \equiv \flow \alpha  = \sum_{i=1}^{\infty}  g^{2i} \gamma_i(\alpha).
\label{eq:gamma}
\ee

Taking the supergauge field into which $A^1_\mu$ and $A^2_\mu$ are embedded
to be $\A_\mu$, and working in the broken phase,
the couplings $g$ and $\alpha$ are
defined through their 
renormalization conditions:
\bea
\label{defg}
	S[\A=A^1]	& =	& {1\over2g^2}\,\str\!\int\!\!d^D\!x\,
									\left(F^1_{\mu\nu}\right)^2+\cdots,
\\	
\label{defg2}
	S[\A=A^2] 	& =	& {1\over2 \alpha g^2}\,\str\!\int\!\!d^D\!x\,
									\left(F^2_{\mu\nu}\right)^2+\cdots,
\eea
where the ellipses stand for higher dimension operators and the
ignored vacuum energy. Note that the renormalization condition
for $g$ constrains the two-point  vertex of
the physical field $S_{\mu \ \; \nu}^{A^1 A^1}(p)$ as follows:
\bea
\label{eq:S_0-11}
	S_{0 \mu  \nu}^{\ 1 \, 1}(p) & = & 2 (p^2 \delta_{\mu\nu} - p_\mu p_\mu) + \Op{4}
\\
\label{eq:S_>0-11}
	S_{n>0 \mu  \nu}^{\ \ \ \ 1 \, 1}(p) & = & \Op{4},
\eea
where we abbreviate $A^1$ by just `1'. 

Defining $\Sigma_i = S_i - 2\hS_i$, the weak coupling flow equations
follow from substituting~\eq{eq:Weak-S}--\eq{eq:gamma}
into the flow equation, as shown in 
\fig{WeakCouplingFE}~\cite{Thesis,mgierg1}.
\bcf[h]
	\be
		\dec{
			\cd{Vertex-n-LdL} 
		}{\{f\}}
		= 
		\dec{
			\begin{array}{c}
				\ds
				\sum_{r=1}^n \left[2 \left(n_r -1 \right) \beta_r +\gamma_r \pder{}{\alpha} \right]\cd{Vertex-n_r-B} 
			\\[4ex]
				\ds
				+ \frac{1}{2} 
				\left( 
					\sum_{r=0}^n \cd{Dumbbell-n_r-r} - \cd{Vertex-Sigma_n_-B} 
				\right)
			\end{array}
		}{\{f\}}
	\label{eq:WeakFlow}
	\ee
\caption{The weak coupling flow equations.}
\label{fig:WeakCouplingFE}
\ecf

We refer to the first two terms on the \rhs\ of~\eq{eq:WeakFlow} as
$\beta$ and $\alpha$-terms, respectively.
The symbol $\bullet$, as in equation~\eq{eq:dot}, means
$-\flowConstAl$. A vertex whose argument is an unadorned letter, say $n$,
represents $S_n$. We define $n_r \equiv n-r$ and $n_\pm = n \pm 1$. The
bar notation of the dumbbell term is defined as follows:
\be
\label{eq:bar}
	a_0[\bar{S}_{n-r}, \bar{S}_r] 	\equiv 	a_0[S_{n-r}, S_r] - a_0[S_{n-r}, \hat{S}_r] - a_0[\hat{S}_{n-r}, S_r].
\ee

The effective propagator relation~\cite{aprop} is central
to the perturbative diagrammatic approach, and arises
from examining the flow of all two-point, tree level vertices.
This is done by setting $n=0$ in~\eq{eq:WeakFlow}
and specialising $\{f\}$ to contain two fields, 
as shown in \fig{TLTPs}.
We note that we can and do choose
all such vertices to be single supertrace terms~\cite{Thesis,mgierg1}.
\bcf[h]
	\be
		\cd{Vertex-TLTP-LdL} = \cd{Dumbbell-S_0-S_0} - \cd{Dumbbell-S_0-hS_0} - \cd{Dumbbell-hS_0-S_0}
	\label{eq:TLTP-flow}
	\ee
\caption{Flow of all possible two-point, tree level vertices.}
\label{fig:TLTPs}
\ecf

Following~\cite{ym,ym1,ym2,aprop,Thesis,mgierg1,scalar2} 
we use the freedom inherent in $\hat{S}$ by choosing the two-point, tree
level seed action vertices equal to the corresponding Wilsonian effective
action vertices. Equation~\eq{eq:TLTP-flow} now simplifies.
Rearranging, integrating \wrt\ $\Lambda$ and choosing the appropriate
integration constants~\cite{Thesis,mgierg1}, we arrive at the
relationship between the integrated \ERG\ kernels---\aka the
effective propagators---and the two-point,
tree level vertices shown in \fig{EPR}. Note
that we have attached the effective propagator, which only
ever appears as an internal line, to an arbitrary structure.
\bcf[h]
	\beas
		\cd{EffPropReln}	& \equiv & \cd{K-Delta} - \cd{FullGaugeRemainder}
	\\
							& \equiv & \cd{K-Delta} - \cd{DecomposedGR}
	\label{eq:EPReln}
	\eeas
\caption{The effective propagator relation.}
\label{fig:EPR}
\ecf

The field
labelled by $M$ can be any of the broken phase fields.
The first term on the \rhs\ is the Kronecker-$\delta$ part of
the effective propagator relation, and the second term is
the gauge remainder part. The gauge remainder decomposes into two
separate components, $\GRk$ and $\GRkpr$, as indicated
on the second line. These individual components will often
be loosely refereed to as gauge remainders
(see~\cite{Thesis,mgierg1,Primer} for
far more detail).

We conclude this section by introducing some new notation.
First, 
we introduce a set of vertex arguments, $v^j$,
where the upper roman index acts as a label. Thus, the $v^j$
are integers, denoting the loop orders of some set of vertices.
Given that both the vertex arguments and number of legs
of the vertices we will shortly encounter are to be summed
over, it is useful to introduce reduced vertices, defined
not to possess a two-point, tree level component. This is
denoted by appending the appropriate vertex
argument with a superscript $R$, \viz\ $v^{j;R}$.

Next, we introduce the compact notation
\beas
	v^{j,j_+} & \equiv & v^j - v^{j+1}
\\
	v^{j,j_+;R} & \equiv & v^{j;R} - v^{j+1;R}.
\eeas
We use this notation to define
\be
\label{eq:CompactVertices}
\SumVertex \equiv \prod_{i=0}^j \sum_{v^{i_+} = 0}^{v^i} \Vertex{v^{i,i_+;R}},
\ee
where the first argument of the structure on the \lhs, $n_s$, gives
the value of $v^0$. Notice that all other vertex arguments are summed over.
The interpretation of the product symbol is as a generator
of  $j+1$ vertices.

The structure shown in~\eq{eq:CompactVertices} always appears as a part of
diagrams which
possess an additional vertex, which carries the argument $v^{j_+}$ (this
argument need not appear on its own---it could be part of something more complicated
\eg $v^{j_+,k}$). An example, which will play an important \role\ later,
is
\be
\label{eq:VertexTower}
	\left[
		\sco[2]{\TopVertex}{\SumVertex}
	\right] \equiv
	\prod_{i=0}^j \sum_{v^{i_+} = 0}^{v^i}
	\left[
	\sco[2]{\TopVertex}{\Vertex{v^{i,i_+;R}}}
	\right].
\ee
Notice that the sum over all vertex arguments is trivially $n_s$:
\be
\label{eq:VertexSum}
	\sum_{i=0}^j v^{i,i_+} + v^{j_+} = \sum_{i=0}^j \left( v^i - v^{i+1} \right) + v^{j+1} = v^0 = n_s.
\ee

The interpretation that the structure defined by~\eq{eq:CompactVertices}
possesses $j+1$ vertices allows us to usefully define~\eq{eq:VertexTower}
for $j=-1$:
\be
\label{eq:Tower-1}
	\left[
		\sco[2]{\TopVertex}{\SumVertex}
	\right]_{j=-1} \equiv
	\Vertex{n^R_s \hspace{0.25em}},
\ee
where $n_s^R$ is, of course, just an $n-s$-loop, reduced vertex. (Note that
this example illustrates the rule that $v^{j_+}$ is replaced by $n$;
this holds irrespective of whether or not $v^{j_+}$ occurs only as part of some
more complicated vertex argument.) We can
even usefully define what we mean by~\eq{eq:VertexTower}
for $j=-2$:
\be
\label{eq:Tower-2}
	\left[
		\sco[2]{\TopVertex}{\SumVertex}
	\right]_{j=-2} \equiv
	\delta(n-s).
\ee
Notice that~\eq{eq:Tower-1} still makes
sense if one and only one of the vertices
is decorated (in practise, we will
always take this to be the top vertex):
if more than one vertex is decorated then this
implies that the number of vertices is at least two,
which leads to a contradiction.
On the other hand, \eq{eq:Tower-2} makes sense
only as is, and not if any of the vertices are
decorated. In the computation of $\beta_n$,
we will find structures like~\eq{eq:VertexTower},
where we sum over $j$. The lower value of this
sum will start out at $-2$. However, as we perform
explicit decorations of the vertices, so we will
need to raise the lower limit on $j$, such
that the diagrams still make sense.

The notation introduced above will allow us to 
conveniently represent the vertices of diagrams
contributing to $\beta_n$. However, diagrams
contain  three additional ingredients: external fields,
effective propagators
and gauge remainder components. We now introduce
notation and rules for these objects, in a manner
compatible with our diagrammatics for the vertices.

To this end, consider the diagrammatic expression
shown in \fig{NewNot}.
\bcf[h]
	\[
	\left[
		\sco[4]{\Vertex{v^{j_+;R}\hspace{0.1em}}}{\Vertex{n_s,j}}
	\right]^{11\Delta^{j+s+1} \GRkpr^m}
	\]
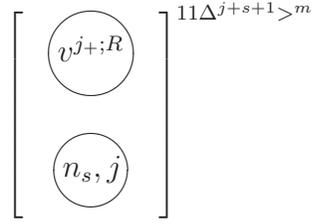
\caption{Representation	 of a set of diagrams in terms of
vertices, external fields, effective propagators and
gauge remainders.}
\label{fig:NewNot}
\ecf

The diagrams represented in \fig{NewNot} possess the following:
\begin{enumerate}
	\item $j+2$ reduced vertices;

	\item two external physical gauge fields, each denoted
		by `1' (Lorentz indices are henceforth suppressed);

	\item $j+s+1$ effective propagators;

	\item $m$ gauge remainder components, $\GRkpr$.
\end{enumerate}

The gauge remainder components behave, diagrammatically, 
in a similar manner to the vertices, in that they form
structures which must be attached to other structures
via internal lines. In this paper, we will not go into
detail about the structures that the gauge remainders
can form (see~\cite{Thesis,InPrep} for more
detail); rather, we note the following. First, each gauge
remainder possesses a `socket' which must be filled by
either an external field or one end of an effective propagator
\ie $\cdeps{Symbol1}$.
Secondly, suppose that we wish to partition $m$ gauge
remainders into two groups of $m'$ and $m-m'$, which form
separate structures. The combinatoric factor associated
with this division is just $\nCr{m'}{m}$, reflecting
the indistinguishability of the $m$ gauge remainders.

Next, we focus on the effective propagators.
Suppose that we wish to join together two vertices
with $q$ effective propagators. First, we note that
each of the $j+2$ vertices is equivalent (before decoration), as can be
straightforwardly checked by a change of variables.
Hence, there are $\nCr{j+2}{2}$ different pairs
of vertices we can chose. Now, we must partition
the effective propagators into two sets containing
$q$ and $j+s+1-q$ elements. There are $\nCr{j+s+1}{q}$
ways to do this. Finally, we note that each effective propagator
can attach with either end to either vertex, yielding a further
factor of $2^q$. Thus, referring to \fig{NewNot}, there are 
\[
	2^q \times \nCr{j+2}{2} \times \nCr{j+s+1}{q}
\]
ways of joining a pair of vertices together
with $q$ effective propagators.
Effective propagators need not join one object to another but
can instead form loops on vertices. In this case, since
the ends of such an effective propagator attach to the same
structure, no factor of two arises from the indistinguishability
of the two ends.

Finally, we can decorate structures with the two external fields.
If the two external fields both attach to the same vertex,
then the combinatoric factor is just unity. If each of the
external fields attaches to a different object then there
is an associated combinatoric factor of two, arising
from the indistinguishability of the two external fields.

The rule for generating the set of 
fully fleshed out diagrams represented in \fig{NewNot}
is simple. First, we form all possible combinations
of gauge remainder structures. Next, we decorate
both the gauge remainder structures and vertices
with the effective propagators and external fields
in all possible ways, ensuring that there are no
empty sockets and that all diagrams are connected.
The combinatorics follows intuitively from the
indistinguishability of the elements of each set
of diagrammatic components from the other
elements in the same set.

\section{An Expression for $\beta_n$}
\label{sec:beta-n}

An expression for $\beta_n$
in terms  of just Wilsonian effective
action vertices, effective propagators
and gauge remainders can be derived by
demonstrating that the set of
diagrams shown in \fig{bn-Pre}, $\LDs{n}$, whose
external momentum we take to be $p$, vanishes at $\Op{2}$.
\bcf[h]
	\bea
	\nonumber
		\ds \LDs{n} & = & \mathcal{D}_n'
							+ 2 \sum_{s=0}^n  \sum_{m=0}^{2s+1} \sum_{j=-2}^{n+s-m-1} 
							\frac{\norm_{j+s+1,j+2}}{m!}
							\dec{
								\dec{
									\sco[1]{
										\TopVertex
									}{\SumVertex}
								}{11\Delta^{j+s+1}\GRkpr^m}
							}{\bullet}
	\\
	\nonumber
					&	& +2 \sum_{s=0}^n  \sum_{m=0}^{2s} \sum_{j=-1}^{n+s-m-2} 
							\frac{\norm_{j+s+1,j+1}}{m!} 
	\\
	\label{eq:beta-n}
					&	& \qquad
							\times \sum_{v^{k}=1}^{v^{j_+}} 
							\left[
								\sco[1]{
									\left[2 \left( v^{j_+,k}-1 \right) \beta_{v^{k}} + \gamma_{v^{k}} \pder{}{\alpha}\right]	
									\Vertex{v^{j_+\!,k}}
								}{\SumVertex}
							\right]	
	\eea
\caption{A set of diagrams which vanishes at $\Op{2}$.}
\label{fig:bn-Pre}
\ecf

$\LDs{n}$ represents a set of $\Lambda$-derivative,
$\beta$ and $\alpha$-terms possessing an $\Op{2}$ stub;
we defer giving an explicit expression for $\LDs{n}'$
until~\cite{InPrep}. 
For the non-negative integers $a$ and $b$,
we define
\be
\label{eq:norm}
	\norm_{a,b} = \frac{(-1)^{b+1}}{a!b!} \left(\frac{1}{2}\right)^{a+1};
\ee
if either $a$ or $b$ is negative, the function is null.

Before moving on, a comment is in order concerning the
ranges of the various summations in \fig{bn-Pre}. 
We begin by 
focusing on the $\Lambda$-derivative 
term. The
value of $s$ controls the sum of the vertex arguments
and thus the maximum loop order vertex that can appear.
If $s$ takes its maximum value of $n$, then the sum
over the vertex arguments is zero: all vertices
are tree level. At the other extreme, the sum over
the vertex arguments is $n$, and so it is clear
that the maximum loop order vertex that can appear is
$n$, as one would expect.

The maximum values of the sums over $m$ and $j$
follow from
the rule that all fully fleshed out diagrams
must be connected. Of the $j+2$ vertices, suppose
that $T$ are tree level vertices. Since the
vertices are reduced (meaning that the tree level
vertices cannot have precisely two decorations), 
and one-point
tree level vertices are automatically zero,
a minimum of $3T$ decorations are required for
the tree level vertices. What of the remaining $j+2-T$
vertices? It is imposed as a constraint that all
one-point, Wilsonian effective action vertices vanish\footnote{
One-point, \emph{seed} action vertices do exist, beyond tree level.
},
in order that the vacuum expectation value of the
superscalar which breaks the $SU(N|N)$ symmetry
is not shifted by quantum corrections~\cite{aprop}. 
This requirement manifests itself as a constraint
on the seed action.
We can thus insist that all Wilsonian effective action vertices are at 
least two-point.
However, it is technically convenient to allow 
the diagrams of \fig{bn-Pre} to possess a single one-point
vertex. If this vertex is struck by $-\flowConstAl$ then it
generates a set of diagrams, not all of which vanish individually.
Using the flow equation to re-express zero in this way essentially
enforces the aforementioned constraint on the seed action `on the
fly'. Hence, we use the following prescription: we allow a single
one-point, Wilsonian effective action vertex \emph{before}
the action of $-\flowConstAl$. After $-\flowConstAl$ has acted, we
adjust the ranges on the sums, as appropriate, to ensure that
no one-point, Wilsonian effective action vertices remain.

Given $T$ tree level vertices, at most one one-point
vertex, and $m$ gauge remainders, we require a minimum of
\be
\label{eq:required}
	2(j+1-T) + 3T + 1 +m = 2j+T+m+3
\ee
decorations. Recalling that effective propagators
are two-ended objects and that we have two
external fields, we see that there are
\[
	2(j+s+1) + 2
\]
available decorations. 
It is thus clear
that 
\be
\label{eq:T+m}
	T+m \leq 2s +1.
\ee
It therefore follows that
\be
\label{eq:Maxm}
	m \leq 2s + 1.
\ee
Next, let us deduce the maximum number of vertices \ie
the maximum number taken by $j+2$ for some values of
$s$ and $m$. From~\eq{eq:VertexSum},
we know that
the sum over vertex arguments is $n-s$.
Therefore, we can have at most $n-s$ vertices which
are not tree level and so a total of $n-s+T$
vertices. Hence,
\[
	j+2 \leq n+s-m+1,
\]
where we have used~\eq{eq:T+m}.

The ranges on the sums for the $\alpha$ and $\beta$
terms follow similarly but now we 
explicitly remove
any diagrams with one-point vertices (since
they cannot be processed) by lowering the upper limits
on the sums over $m$ and $j$ by one apiece. Furthermore, the lower
limit of the sum over $j$ is $-1$ and not $-2$,
since there must be at least one vertex present.

Having justified the ranges on the various sums,
we proceed by simply allowing $-\flowConstAl$ to act.
In the case where
$-\flowConstAl$ strikes a vertex (in which case
we must take the lower limit of the sum over $j$
to be $-1$), the 
$\beta$ and $\alpha$-terms thus generated exactly cancel 
the $\beta$ and $\alpha$-terms
in \fig{bn-Pre}. If $-\flowConstAl$ strikes
either an effective propagator or a gauge remainder,
we adjust the ranges on the sums over $m$ and $j$
to remove
any Wilsonian effective action one-point vertices.
Having done this, we can increase the lower limit
on $s$ to remove certain diagrams which vanish at $\Op{2}$.
This works as follows. Suppose that $s=0$ and
that all diagrams with one-point vertices have been
discarded. There are $2j+4$ fields available
to decorate the $j+2$ vertices and so each vertex
must be exactly two-point. 
However, recall from~\eq{eq:S_>0-11}
that the $\Op{2}$ part of
all $n>0$ loop, two-point vertices vanishes.
As with diagrams possessing one-point vertices,
it is useful to keep diagrams known to vanish at
$\Op{2}$ until after $-\flowConstAl$ has acted. The
resulting terms which individually possess $\Op{2}$
components are all involved in diagrammatic cancellations.

Both $a_0$ and $a_1$ have the capacity
to  replace a
one-point Wilsonian effective action vertex
with something else. In the former case,
a one-point Wilsonian effective action
vertex is replaced by a dumbbell structure
possessing a one-point vertex;  the component
of this structure in which the one-point 
vertex is a seed action vertex survives. In the
latter case, a one-point, Wilsonian effective
action vertex is replaced by a padlock
structure which is decorated by a single field.

Now let us consider the effect of the action
of $a_1$ in more detail.
Due to the equivalence of each of the vertices
in a given term, we can take the $a_1$ to strike
the vertex with argument $v^{j_+;R}$ (so long as we multiply by $j+2$), causing
the argument to become
$\Sigma_{v^{j_+}-1}$.
The $R$ has been dropped,
since a quantum term is necessarily formed from a 
vertex whose argument is greater than zero. The
lower limit on the sum over $v^{j_+}$ should now
be changed from zero to one, in recognition of the
fact that
$a_1$ does not act on tree level terms. Furthermore,
the sum over all vertex arguments is now $n+s-1$,
rather than $n+s$, and so we should reduce the
upper limit of the sum over $s$ by one. It is
convenient to change variables:
\beas
	v^{j_+}	& \rightarrow	& v^{j_+} + 1
\\	
	v^j		& \rightarrow	& v^j +1
\\
			& \vdots		&
\eeas
and then to let $s \rightarrow s-1$. 

The action of $a_0$ requires care since
we must take the reduction of a
vertex struck by $a_0$ seriously.
This follows because terms formed by $a_0$,
 unlike $\alpha$-terms, $\beta$-terms
or terms formed 
by $a_1$, 
can be spawned by
tree level vertices.
Recall 
that a reduced vertex lacks a two-point,
tree level component.
The flow of a reduced vertex
must therefore lack a component given by 
the flow of a two-point, tree level vertex.
It follows from~\eq{eq:TLTP-flow}
that the 
dumbbell structure
generated by the flow of a reduced vertex
either possesses at least one reduced vertex or the
kernel is decorated. We denote the reduction
of a dumbbell structure by tagging the structure with
$R$. The terms generated by allowing $-\flowConstAl$ to act
in \eq{eq:beta-n} are shown in \fig{bn-P}.
\bcf[h]
	\[
	\begin{array}{c}
	\vspace{4ex}
		\ds
		\LDs{n} \rightarrow \LDs{n}' - \sum_{s=0}^n   \sum_{m=0}^{2s+1} \sum_{j=-1}^{n+s-m-1}\frac{\norm_{j+s+1,j+1}}{m!}
		\ds
				\dec{
					\sco[1]{
						\ds
						\sum_{v^{k}=1}^{v^{j_+}}
						\left[
							\LDi{Dumbbell-vj_+kb-vkb}{D-vj_+kb-vkb}
						\right]_R \hspace{-2pt}
					}{\Vertex{ n_s, j}}
				}{11\Delta^{j+s+1} \GRkpr^m}
	\\
	\vspace{4ex}
		\ds
		+\sum_{s=1}^n   \sum_{m=0}^{2s-1} \sum_{j=-1}^{n+s-m-2}
		\frac{\norm_{j+s,j+1}}{m!}
		\dec{
			\sco[1]{
				\LDi{Padlock-Sig_vj_+}{P-Sig_vj_+}
			}{\Vertex{ n_{s}, j}}
		}{11\Delta^{j+s} \GRkpr^m}
	\\
	\vspace{4ex}
		\ds
		+ \sum_{s=1}^n   \sum_{m=0}^{2s} \sum_{j=-2}^{n+s-m-2}
		\frac{\norm_{j+s,j+2}}{m!}
		\left[
			\dec{
				\sco[1]{\LO{\Vertex{v^{j_+\!;R}}}{j+2-DEP}}{\Vertex{n_s, j}}
			}{11\Delta^{j+s} \DEP \GRkpr^m}
		\right]
	\\
		\ds
		+2 \sum_{s=1}^n   \sum_{m=1}^{2s} \sum_{j=-2}^{n+s-m-2}
		\frac{\norm_{j+s+1,j+2}}{(m-1)!}
		\dec{
			\sco[1]{\LO{\Vertex{v^{j_+\!;R}}}{j+2-DGR}}{\Vertex{n_s, j}}
		}{11\Delta^{j+s+1} \GRkpr^{m-1} \DGRkpr}
	\end{array}
	\]
\caption{Expression for $\LDs{n}$ obtained by allowing
$-\flowConstAl$ to act in~\eq{eq:beta-n} and discarding
terms which vanish at $\Op{2}$ or possess a one-point,
Wilsonian effective action vertex.}
\label{fig:bn-P}
\ecf

We now isolate all two-point, tree level
vertices in diagram~\ref{D-vj_+kb-vkb}.
This step is crucial to
the entire diagrammatic procedure: if a
two-point, tree level vertex is attached to an effective propagator,
we can employ the effective propagator relation, whereas
if it is attached to an external field, we can perform
manipulations at $\Op{2}$.

Let us consider isolating the two-point, tree level
vertices of diagram~\ref{D-vj_+kb-vkb} in more detail.
First, let us take both vertices
of the dumbbell structure to be reduced vertices.
This immediately allows us to reduce the maximum 
value of the sum over $j$ by one, since the
total number of reduced vertices is now $j+3$,
rather than $j+2$. Of the terms which remain, consider
those possessing exclusively 
Wilsonian 
effective action vertices. Since we discard
one-point, Wilsonian effective action vertices
in all diagrams in which $\flowConstAl$ has acted,
we can reduce the maximum values of 
$j$ and $m$ and increase
the minimum value of $s$ by one.
The component of the resulting diagram in which
the kernel is undecorated is exactly cancelled
by the component of diagram~\ref{j+2-DEP}
in which the differentiated effective propagator
joins two different vertices. Since the surviving
component has a decorated kernel,
the maximum values of both $m$ and $j$ are reduced
by one, again.

Isolating a single, two-point, tree level vertex
in diagram~\ref{D-vj_+kb-vkb} is straightforward:
taking the argument of the top or bottom vertex
of the dumbbell structure to be a two-point, tree level
vertex amounts to the same thing; hence we will choose
to isolate the two-point, tree level part of $\bar{v}^k$ 
and multiply by two.
When taking the tree level
part of $\bar{v}^k$,
the other vertex argument,
$\bar{v}^{j_+,k;R} \equiv \bar{v}^{j_+;R} - \bar{v}^{k;R}$, reduces to 
simply $\bar{v}^{j_+;R}$. The equality of the two-point, tree
level Wilsonian effective action vertices allows us
to simplify the bar notation: from~\eq{eq:bar}
only the seed action component of $\bar{v}^{j_+;R}$ (which comes
with a minus sign) survives. There is no need to change the 
limits on any of the sums: compared to the
parent diagram, we have an extra two-point vertex,
but also an extra two decorative fields, corresponding
to the two ends of the kernel.

Taking two two-point, tree level vertices in
diagram~\ref{D-vj_+kb-vkb} requires some thought.
First, we note that by the definition of $R$
acting on a dumbbell, the kernel must be decorated,
which we indicate by the notation
\DecKernel.
Secondly, the dumbbell
structure cannot have been formed by
the flow of a one-point vertex. Since we are
interested only in one-point Wilsonian effective
action vertices if
they have been processed,
we can reduce the maximum values of $m$
and $j$ and increase the
minimum value of $s$ to remove any unwanted diagrams.
If $j=-1$, then there are only two vertices
in total, and so this case should be treated differently
from $j \geq 0$. In the latter case, it will prove
useful to shift variables $j \rightarrow j+1$, so
that the sum over $j$ starts, once again, from $-1$.
The isolation of the
two-point, tree level vertices of 
diagram~\ref{D-vj_+kb-vkb}, together with
the cancellation of the component of diagram~\ref{j+2-DEP} in which the
differentiated effective propagator joins two separate vertices, is shown in 
\fig{bn-Ia}. 
\bcf[h]
	\[
		\begin{array}{c}
		\vspace{2ex}
			-
					\ds
					\sum_{s=1}^n \sum_{m=0}^{2s-1}  \sum_{j=-1}^{n+s-m-4} 
					\frac{\norm_{j+s+1,j+1}}{m!}
					\sum_{v^{k}=1}^{v^{j_+}} 
					\dec{
						\LO{\cd{Dumbbell-vj_+kR-RW-vkR} \Vertex{n_s,j}}{D-vj_+kR-vkR}
					}{11\Delta^{j+s+1} \GRkpr^m}
		\\
		\vspace{2ex}
			+
			\left[
				\begin{array}{c}
				\vspace{1ex}
					\ds
					\sum_{m=0}^{2n-1} \frac{\norm_{n,0}}{m!} 
				\\
					\dec{
						\LDi{Dumbbell-02x2-R}{D-02x2}
					}{11\Delta^n\GRkpr^m}
				\end{array}
			\right]
			+2
			\left[
				\begin{array}{c}
				\vspace{1ex}
					\ds
					\sum_{s=0}^n \sum_{m=0}^{2s+1}  \sum_{j=-1}^{n+s-m-2} 
					\frac{\norm_{j+s+1,j+1}}{m!}
				\\
					\ds
					\sum_{v^{k}=1}^{v^{j_+}} 
					\dec{
						\LO{\cd{Dumbbell-vj_+kR-vkhR} \Vertex{n_s,j}}{D-vj_+kR-vkhR}
					}{11\Delta^{j+s+1} \GRkpr^m}
				\end{array}
			\right]
		\\
			+
			\left[
				\begin{array}{c}
				\vspace{1ex}
					\ds
					\sum_{s=1}^n \sum_{m=0}^{2s-1}  \sum_{j=-1}^{n+s-m-3} 
					\frac{\norm_{j+s+2,j+2}}{m!}
				\\
					\ds
					\dec{
						\begin{array}{c}
						\vspace{1ex}
							\ds
							\LDi{Dumbbell-02x2-R}{D-02x2-R-VS}
						\\
						\vspace{1ex}
							\Vertex{v^{j_+\!;R}}
						\\
							\Vertex{n_s, j}
						\end{array}
					}{11\Delta^{j+s+2} \GRkpr^m}
				\end{array}
			\right]
			+2
			\left[
				\begin{array}{c}
				\vspace{1ex}
					\ds
					\sum_{s=0}^n \sum_{m=0}^{2s+1}  \sum_{j=-1}^{n+s-m-1} 
					\frac{\norm_{j+s+1,j+1}}{m!}
				\\
					\ds
					\dec{
						\sco[1]{
							\ds				
							\LDi{Dumbbell-vj_+h-02}{D-vj_+R-02}
						}{\Vertex{n_s, j}}		
					}{11\Delta^{j+s+1} \GRkpr^m}	
				\end{array}
			\right]
		\end{array}
	\]
\caption{Isolation of the two-point, tree level vertices
of diagram~\ref{D-vj_+kb-vkb}. The effect of cancelling
the component of diagram~\ref{j+2-DEP} in which the
differentiated effective propagator joins two separate vertices
forces the kernel of diagram~\ref{D-vj_+kR-vkR} to possess
only decorated components.}
\label{fig:bn-Ia}
\ecf

The next step of the diagrammatic procedure
is to decorate the two-point, tree level 
vertices of diagrams~\ref{D-02x2}, \ref{D-02x2-R-VS} and~\ref{D-vj_+R-02},
with either
an external field
or an end of an effective propagator. 
In the latter case we must then attach
the loose end of any effective propagators
to an available structure. 
We refer to the primary part of such 
a diagram as the component left over after
applying the effective propagator relation
as many times as possible but, each time,
retaining only the 
Kronecker-$\delta$ contribution.

Assuming that
the necessary structures exist, we can do the
following with a diagram possessing a single
two-point, tree level vertex:
\begin{enumerate}
	\item	attach an external field;

	\item	attach one end of an effective propagator,
			with the other end attaching to:
			\begin{enumerate}
				\item	one of the Wilsonian  effective action vertices;
				\label{EP-DV}

				\item	the seed action vertex to which the kernel
						attaches;
				\label{EP-SV}

				\item	the kernel;
				\label{EPx2-V-k}

				\item	a gauge remainder.
				\label{EP-GR}
			\end{enumerate}
\end{enumerate}

In each of~\ref{EP-DV}--\ref{EP-GR}
the effective propagator relation
can be applied. For the purposes of
this paper, we will concern ourselves
with just the primary part
coming from~\ref{EP-DV} and~\ref{EP-SV};
the treatment of the
gauge remainder contributions and~\ref{EP-GR} 
is deferred until~\cite{InPrep}.\footnote{The
primary part coming from~\ref{EPx2-V-k}
corresponds to a kernel which bites its
own tail, and so can be discarded.};

\begin{cancel}[Diagam~\ref{D-vj_+kR-vkhR}]
\label{Cancel:D-vj_+kR-vkhR}
Consider the primary part of
diagram~\ref{D-vj_+R-02} corresponding
to~\ref{EP-DV}, above, which we note
exists only for $j>-1$. 
For comparison with diagram~\ref{D-vj_+kR-vkhR},
it is convenient
to change variables $j \rightarrow j+1$,
so that the sum over $j$ once again starts
from $-1$ and
to identify $\hat{v}^{j+2}$ with $\hat{v}^{k}$.
Thus, 
the two-point, tree level vertex
can be joined to any of $j+2$ identical Wilsonian
effective action vertices, using
any of $j+s+2$ effective propagators.
Noting that this effective propagator can
attach either way round,  the
combinatoric factor is 
\[
	2 (j+s+2)(j+2),
\]
which, from \eq{eq:norm}, 
combines with $\norm_{j+s+2,j+2}$
to give $-\norm_{j+s+1,j+1}$. Thus,
the primary part of
diagram~\ref{D-vj_+R-02} corresponding
to~\ref{EP-DV}, above,
precisely cancels
diagram~\ref{D-vj_+kR-vkhR}.
\end{cancel}

Before discussing cancellations arising
from the primary part of diagram~\ref{D-vj_+R-02} corresponding
to~\ref{EP-SV}, we consider a cancellation
arising from one of the diagrams possessing
two two-point vertices.

\begin{cancel}[Diagram~\ref{D-vj_+kR-vkR}]
Consider attaching each two-point, tree
level vertex of diagram~\ref{D-02x2-R-VS}
to a separate reduced vertex. 
The primary part of this diagram
precisely cancels
diagram~\ref{D-vj_+kR-vkR}.
\end{cancel}

Similarly, it is straightforward to show that
all components of diagram~\ref{P-Sig_vj_+}
are cancelled by the following terms:
\begin{enumerate}
	\item 	The component of diagram~\ref{j+2-DEP}
			in which the differentiated effective
			propagator attaches at both ends to the
			same vertex;

	\item	The primary part of diagram~\ref{D-vj_+R-02}
			corresponding to~\ref{EP-SV}, above;

	\item	The primary part of diagram~\ref{D-02x2-R-VS}
			in which both two-point, tree level vertices
			are attached to the same reduced vertex;

	\item	The primary parts of diagrams~\ref{D-02x2} and~\ref{D-02x2-R-VS}
			in which the two-point, tree level vertices are
			attached to each other.

\end{enumerate}

We have thus demonstrated that, up to terms involving gauge remainders
or possessing an $\Op{2}$ stub, $\LDs{n}$ does indeed vanish
at $\Op{4}$. The complete proof, to be given in~\cite{InPrep},
is considerably more arduous than
the partial demonstration given here, requiring further diagrammatic techniques.
However, given that $\LDs{n}$ does vanish at $\Op{4}$ we can then
directly derive an expression for $\beta_n$ in terms of just
Wilsonian effective action vertices, effective propagators
and (components of) gauge remainders.
First, we specialize
$\LDs{n}$ to $\Op{2}$ and so 
discard
all terms which are manifestly $\Op{4}$. Then we notice
that the only remaining term containing $\beta_n$ multiplies
a two-point, tree level vertex. However, from~\eq{eq:S_0-11},
the $\Op{2}$ part
of this vertex is just a universal coefficient. Pulling
the $\beta_n$ term to the other side of the equation, we have an expression
for $\beta_n$ in terms of 
Wilsonian effective action vertices, effective propagators,
(components of) gauge remainders
and $\beta_{m<n}$. These latter
terms can be substituted for using the diagrammatic expression
for $\beta_{m<n}$ and so forth, until all $\beta$ terms
have been re-expressed diagrammatically.

\end{document}